\documentclass[aps,preprint]{revtex4}%
\usepackage{amsfonts}
\usepackage{amsmath}
\usepackage{amssymb}
\usepackage{graphicx}%
\providecommand{\U}[1]{\protect\rule{.1in}{.1in}}
\newtheorem{theorem}{Theorem}

\newtheorem{conjecture}[theorem]{Conjecture}

\begin{document}
\preprint{ }
\title[ ]{SU(5) Grand Unified Model and Dark Matter}
\author{Shi-Hao Chen}
\affiliation{Institute of Theoretical Physics, Northeast Normal University, Changchun
130024, P.R.China; shchen@nenu.edu.cn}

\begin{abstract}
A dark matter model which is called w-matter or mirror dark matter is
concretely constructed based on (f-SU(5))X(w-SU(5)) symmetry. There is no
Higgs field and all masses originate from interactions in the present model.
W-matter is dark matter relatively to f-matter and vice versa. In high-energy
processes or when temperature is very high, visible matter and dark matter can
transform from one into another. In such process energy seems to be
non-conservational, because dark matter cannot be detected. In low-energy
processes or when temperature is low, there is only gravitation interaction of
dark matter for visible matter.

\end{abstract}
\volumeyear{year}
\volumenumber{number}
\issuenumber{number}
\eid{identifier}
\date[Date text]{date}
\received[Received text]{date}

\revised[Revised text]{date}

\accepted[Accepted text]{date}

\published[Published text]{date}

\startpage{1}
\endpage{7}
\maketitle
\tableofcontents

\section{Introduction}

What is the origin of mass? A possible answer is spontaneous
symmetry-breaking. Higgs fields can cause spontaneous symmetry-breaking. But
it is difficult to understand $\left(  -\mu^{2}\right)  $ in Higgs potentials.
Hence dynamical breaking is considered$^{[1]}$. It is not realized to
construct a realistic grand unified model based the dynamical breaking
according to the conventional theory.

There are many sorts of grand unified models. There are some difficulties such
as proton decay in the simple $SU(5)$ model. There are not the proton decay
and quark confinement problems in a $SU(5)$ model with hadrons as
nontopological solitons$^{[2]}.$ This model is not contradictory to given
experiments and astronomical observations up to now. Hence a $SU(5)$ model is
still possible.

Many astronomical observations show that there is dark matter. Many dark
matter models were presented. A necessary inference of a quantum field theory
without divergence is just that there must be dark matter ($w-matter$) which
and visible matter are symmetric and there is no interaction except the
gravitation between both$^{[3]}$. The energy density $\rho_{0}$ is zero
without normal ordering of operators and all loop corrections are finite in
the quantum field theory. The sort of dark matter ($w-matter$) is called
mirror matter which is discussed in detail in Refs$^{\left[  4\right]  }$.

A dark matter model which is called $w-matter$ or mirror dark matter is
concretely constructed based on $SU_{f}(5)\times SU_{w}(5)$ symmetry in the
present paper. There is no Higgs field and all masses originate from
interactions in the present model. $W-matter$ is dark matter relatively to
$f-matter$ and vice versa. In high-energy processes or when temperature is
very high, visible matter and dark matter can transform from one into another.
In such process energy seems to be non-conservational, because dark matter
cannot be detected. In low-energy processes or when temperature is low, there
is only gravitation interaction of dark matter for visible matter.

In section 2, Lagrangian of the $SU_{f}(5)\times SU_{w}(5)$ model is
constructed; In section 3, symmetry spontaneously breaking is discussed; In
section 4, the physical significance of the present model is given; Section 5
is the conclusion.

\section{Lagrangian of the $SU_{f}(5)\times SU_{w}(5)$ model}

\begin{conjecture}
There are two sorts of matter which are called $fire-matter$ ($f-matter$) and
$water-matter$ ($w-matter$), respectively. Both are symmetric and have
$SU_{f}(5)\times SU_{w}(5)$ symmetry. There is no other interaction except the
gravitation between both and the coupling $\left(  5\right)  $ of f-scalar
fields and w-scalar fields.
\end{conjecture}

The conjecture, in fact, is a necessary inference of a quantum field theory
without divergence in which all loop-corrections are finite and the energy
density $\rho_{0}$ of the vacuum state must be zero without normal ordering of
operators$^{\left[  3\right]  }$. It is obvious that the conjecture is
consistent with a sort of dark matter model which is called $w-matter^{\left[
3\right]  }$ or mirror dark matter$^{\left[  4\right]  }$.

Based the conjecture, the Lagrangian density of the $SU_{f}(5)\times
SU_{w}(5)$ model can be taken as
\begin{align}%
\mathcal{L}%
&  =%
\mathcal{L}%
_{f}\left(  \chi_{f},\Psi_{f},G_{f},\Phi_{f},H_{f}\right)  +%
\mathcal{L}%
_{w}\left(  \chi_{w},\Psi_{w},G_{w},\Phi_{w},H_{w}\right)  +%
\mathcal{L}%
_{\Omega}+V,\tag{1}\\
V  &  =V_{f}+V_{w}+V_{\Omega}+V_{I},\nonumber
\end{align}%
\begin{equation}
V_{f}=\frac{1}{4}a\left(  Tr\Phi_{f}^{2}\right)  ^{2}+\frac{1}{2}bTr\left(
\Phi_{f}^{4}\right)  +\frac{1}{4}\xi\left(  H_{f}^{+}H_{f}\right)  ^{2}%
+\frac{1}{2}\varsigma H_{f}^{+}H_{f}Tr\Phi_{f}^{2}-\frac{1}{2}\varkappa
H_{f}^{+}\Phi_{f}^{2}H_{f}, \tag{2}%
\end{equation}%
\begin{equation}
V_{w}=\frac{1}{4}a\left(  Tr\Phi_{w}^{2}\right)  ^{2}+\frac{1}{2}bTr\left(
\Phi_{w}^{4}\right)  +\frac{1}{4}\xi\left(  H_{w}^{+}H_{w}\right)  ^{2}%
+\frac{1}{2}\varsigma H_{w}^{+}H_{w}Tr\Phi_{w}^{2}-\frac{1}{2}\varkappa
H_{w}^{+}\Phi_{w}^{2}H_{w},, \tag{3}%
\end{equation}%
\begin{equation}
V_{\Omega}=\frac{1}{4}\lambda\Omega^{4},\text{ \ \ }%
\mathcal{L}%
_{\Omega}=\frac{1}{2}\partial_{\mu}\Omega\partial^{\mu}\Omega, \tag{4}%
\end{equation}%
\begin{equation}
V_{I}=-\frac{1}{15}w\Omega^{2}\left(  Tr\Phi_{f}^{2}+Tr\Phi_{w}^{2}\right)
-\frac{2A}{225}Tr\Phi_{f}^{2}Tr\Phi_{w}^{2}, \tag{5}%
\end{equation}
where $\chi$ and $\Psi$ denote fermion fields, and $G$ the $SU(5)$ gauge
fields. $\Omega,$ $\Phi$\ and $H$\ are\ the $\underline{1},$\ $\underline{24}
$ and $\underline{5}$ representations, respectively. It should be pointed out
that all the scalar fields are not Higgs fields because they are all massless
before symmetry breaking.

Similarly to the conventional $SU(5)$ model, the possible fermion states for
the first generation are
\begin{equation}
\Psi_{fL}=\frac{1}{\sqrt{2}}\left(
\begin{array}
[c]{c}%
0\text{ \ \ \ \ }u_{f3}^{c}\text{\ }-u_{f2}^{c}\text{ }-u_{f1}\text{ }%
-d_{f1}\\
-u_{f3}^{c}\text{ \ \ }0\text{ \ \ \ }u_{f1}^{c}\text{\ }-u_{f2}%
\text{\ }-d_{f2}\\
u_{f2}^{c}\text{ }-u_{f2}^{c}\text{ \ \ }0\text{ \ }-u_{f3}\text{\ }-d_{f3}\\
u_{f1}\text{ \ \ \ }u_{f2}\text{ \ \ }u_{f3}\text{ \ \ \ }0\text{ }-e_{f}%
^{+}\\
d_{f1}\text{ \ \ }d_{f2}\text{ \ \ }d_{f3}\text{ \ \ \ }e_{f}^{+}\text{ \ \ }0
\end{array}
\right)  _{L},\text{ \ }\Psi_{fR}=\left(
\begin{array}
[c]{c}%
d_{f1}\\
d_{f1}\\
d_{f1}\\
e_{f}^{+}\\
-\nu_{fe}^{c}%
\end{array}
\right)  _{R} \tag{6}%
\end{equation}%
\begin{equation}
\Psi_{wR}=\frac{1}{\sqrt{2}}\left(
\begin{array}
[c]{c}%
0\text{ \ \ \ \ \ \ }u_{w3}^{c}\text{\ \ }-u_{w2}^{c}\text{ }-u_{w1}\text{
}-d_{w1}\\
-u_{w3}^{c}\text{ \ \ \ }0\text{ \ \ \ \ }u_{w1}^{c}\text{\ \ }-u_{w2}%
\text{\ }-d_{w2}\\
u_{w2}^{c}\text{ \ }-u_{w2}^{c}\text{ \ \ }0\text{ \ \ }-u_{w3}\text{\ }%
-d_{w3}\\
u_{w1}\text{ \ \ \ }u_{w2}\text{ \ \ }u_{w3}\text{ \ \ \ }0\text{ }-e_{w}%
^{+}\\
d_{w1}\text{ \ \ }d_{w2}\text{ \ \ }d_{w3}\text{ \ \ \ }e_{w}^{+}\text{ \ \ }0
\end{array}
\right)  _{R},\text{ \ }\Psi_{wL}=\left(
\begin{array}
[c]{c}%
d_{w1}\\
d_{w1}\\
d_{w1}\\
e_{w}^{+}\\
-\nu_{we}^{c}%
\end{array}
\right)  _{L} \tag{7}%
\end{equation}
The other possible model is an $SU(5)$ grand unified model with hadrons as
nontopological solitons$^{[2]}.$ The conclusions of the present paper are
independent of a concrete model.

\section{Symmetry spontaneously breaking and temperature effects}

For simplicity, we do not consider the couplings $\Omega$ and $\Phi$ with
$\chi$ for a time. Ignoring the contributions of the scalar fields and the
fermion fields to one loop correction and only considering the contribution of
the gauge fields to one-loop correction, when $\overline{\varphi}_{s}\ll kT$,
$k$ is the Boltzmann constant, similarly to Ref. $[1],$ the finite-temperature
effective potential approximate to 1-loop in flat space can be obtained
\begin{align}
V  &  =\frac{\lambda}{8}T^{2}\Omega^{2}+\frac{1}{4}\lambda\Omega^{4}-\frac
{A}{2}\varphi_{f}^{2}\varphi_{w}^{2}-\frac{1}{2}w\Omega^{2}\left(  \varphi
_{f}^{2}+\varphi_{w}^{2}\right) \nonumber\\
&  +\frac{D}{4!}\varphi_{f}^{4}+B\varphi_{f}^{4}\left(  \ln\frac{\varphi
_{f}^{2}}{\sigma^{2}}-\frac{1}{2}\right)  +CT^{2}\varphi_{f}^{2}\nonumber\\
&  +\frac{D}{4!}\varphi_{w}^{4}+B\varphi_{w}^{4}\left(  \ln\frac{\varphi
_{w}^{2}}{\sigma^{2}}-\frac{1}{2}\right)  +CT^{2}\varphi_{w}^{2}, \tag{8}%
\end{align}
where
\begin{equation}
\Phi_{s}=Diagonal\left(  1,1,1,-\frac{3}{2},-\frac{3}{2}\right)
\overline{\varphi}_{s}, \tag{9}%
\end{equation}%
\[
B\equiv\frac{5625}{1024\pi^{2}}g^{4},\text{ \ }\frac{\left(  225a+105b\right)
}{16}\equiv\frac{D}{4!}+\frac{11}{3}B,\text{ \ }C\equiv\frac{75}{16}\left(
kg\right)  ^{2},
\]
$\sigma$ is regarded as a constant, and the terms independent of $\Omega$ and
$\Phi$ are neglected.

According to the mirror dark matter model, the temperature of mirror matter is
strikingly lower than that of visible matter. But this is not necessary when a
cosmological model is considered. We will discuss the problem in another
paper. The temperature $T_{f}$ of $f-matter$ may be different from $T_{w}$ of
$w-matter$ in the present model as well, but for simplicity we take
$T_{f}=T_{w}.$

The conditions by which $V$ takes its extreme values are
\begin{align}
\left[  \lambda\overline{\Omega}^{2}-w\left(  \overline{\varphi}_{f}%
^{2}+\overline{\varphi}_{w}^{2}\right)  +\frac{\lambda}{4}T^{2}\right]
\overline{\Omega}  &  =0,\tag{10a}\\
-w\overline{\Omega}^{2}-A\overline{\varphi}_{w}^{2}+\frac{D}{6}\overline
{\varphi}_{f}^{2}+4B\overline{\varphi}_{f}^{2}\ln\frac{\overline{\varphi}%
_{f}^{2}}{\sigma^{2}}+2CT^{2}  &  =0,\tag{10b}\\
-w\overline{\Omega}^{2}-A\overline{\varphi}_{f}^{2}+\frac{D}{6}\overline
{\varphi}_{w}^{2}+4B\overline{\varphi}_{w}^{2}\ln\frac{\overline{\varphi}%
_{w}^{2}}{\sigma^{2}}+2CT^{2}  &  =0. \tag{10c}%
\end{align}
When $T\sim0$,
\begin{align}
\overline{\varphi}_{f}^{2}  &  =\overline{\varphi}_{w}^{2}\equiv\sigma_{0}%
^{2}=\sigma^{2}\exp M,\text{ \ \ }M\equiv\frac{1}{4B}\left(  A+\frac{2w^{2}%
}{\lambda}-\frac{D}{6}\right)  ,\nonumber\\
\overline{\Omega}_{0}^{2}  &  =\upsilon_{0}^{2}=\frac{2w}{\lambda}\sigma
^{2}\exp M,\tag{11a}\\
V  &  =V_{\min}=-B\sigma^{4}\exp2M. \tag{11b}%
\end{align}
$\sigma^{2}\left(  T\right)  $ and $\upsilon^{2}\left(  T\right)  $ will
decrease and $V_{\min}$ will increase as temperature rises. There must be the
critical temperature $T_{cr}$ so that when $T>T_{cr},$ the least value of $V$
is $V\left(  \overline{\varphi}_{f}=\overline{\varphi}_{w}=\overline{\Omega
}=0\right)  =0.$ $T_{cr}$ is rough estimated to be
\begin{equation}
T_{cr}=\frac{8B}{w+8C}\sigma^{2}\exp\left(  M-\frac{1}{2}\right)  . \tag{12}%
\end{equation}

$\Omega$ is not absolutely necessary for the symmetry breaking of the present
model, but it is necessary for some a cosmological model$^{[5]}.$

After spontaneous symmetry-breaking, the reserved symmetry is $\left[
SU_{f}(3)\times SU_{f}(2)\times U_{f}(1)\right]  \times\left[  SU_{w}(3)\times
SU_{w}(2)\times U_{w}(1)\right]  .$ The breaking is a sort of dynamical
breaking. In other words, the interactions of the scalar fields with the gauge
fields make the massless scalar fields become `Higgs fields', and finally
cause the spontaneous symmetry-breaking. As a consequence, the $f-particles$
($w-particles$) can get their masses determined by the reserved symmetry
$SU(3)\times SU(2)\times U(1)$ as the conventional $SU(5)$ $GUT$ theory in
which there are Higgs fields.

\section{The physical significance of the present model}

1. The model implies that all masses originate from interactions.

2. $W-matter$ is dark matter for $f-matter$ in low energy process, vice versa.
This is because the masses of the scalar particles to be very large in low
temperature so that the transformation of the $f-$ and the $w-scalar$
particles from one into another and their effects may be ignored and there is
no interaction except the coupling $\left(  5\right)  $ and the gravitation
between $f-matter$ and $w-matter$. This sort of dark matter is called mirror
dark matter in Refs.$^{\left[  4\right]  }$.

3. In high-energy processes or when temperature is very high, visible matter
and dark matter can transform from one into another. In such process energy
seems to be non-conservational, because dark matter cannot be detected. The
following reaction originating from $\left(  1\right)  $ and $\left(
5\right)  $ is an example in which visible matter transforms into dark matter.%
\begin{equation}
p+\overline{p}\longrightarrow\varphi_{fA}\longrightarrow\varphi_{fB}%
+\varphi_{wC}+\varphi_{wD}. \tag{13}%
\end{equation}
In the reaction $\varphi_{wC}$ and $\varphi_{wD}$ and the $w-particles$ coming
from the decay of $\varphi_{wC}$ and $\varphi_{wD}$ cannot be detected.

\section{Conclusion}

A dark matter model which is called $w-matter$ or mirror dark matter is
concretely constructed based on $SU_{f}(5)\times SU_{w}(5)$ symmetry. There is
no Higgs field and all masses originate from interactions in the present
model. $W-matter$ is dark matter relatively to $f-matter$ and vice versa. In
high-energy processes or when temperature is very high, visible matter and
dark matter can transform from one into another. In such process energy seems
to be non-conservational, because dark matter cannot be detected. In
low-energy processes or when temperature is low, there is only gravitation
interaction of dark matter for visible matter.

\textbf{Acknowledgement}

I am very grateful to professor Zhao Zhan-yue and professor Wu Zhao-yan for
their helpful discussions and best support.

\end{document}